# Non-Covalent Dimerization after Enediyne Cyclization on Au(111)

*Dimas G. de Oteyza,*[*,‡,1,2] *Alejandro Pérez Paz,* [‡,3] *Yen-Chia Chen,*[4] *Zahra Pedramrazi,*[4] *Alexander Riss,*[4] *Sebastian Wickenburg,*[4] *Hsin-Zon Tsai,*[4] *Felix R. Fischer,*[5,6,7] *Michael F. Crommie,* [4,6,7] *Angel Rubio* [3,8,9]

[1] Donostia International Physics Center, E-20018 San Sebastián, Spain

[2] Ikerbasque, Basque Foundation for Science, E-48011 Bilbao, Spain

[3] Nano-Bio Spectroscopy Group and ETSF, Universidad del País Vasco, CFM CSIC-UPV/EHU-MPC, 20018 San Sebastián, Spain

[4] Department of Physics, University of California, Berkeley, CA 94720, USA

[5] Department of Chemistry, University of California, Berkeley, CA 94720, USA

[6] Materials Sciences Division, Lawrence Berkeley National Laboratory, Berkeley, CA 94720, USA

[7] Kavli Energy NanoSciences Institute at the University of California Berkeley and the Lawrence Berkeley National Laboratory, Berkeley, CA 94720, USA

[8] Max Planck Institute for the Structure and Dynamics of Matter, Luruper Chaussee 149, 22761 Hamburg, Germany

[9] Center for Free-electron Laser Science (CFEL), Luruper Chaussee 149, 22761 Hamburg, Germany

**ABSTRACT:** We investigate the thermally-induced cyclization of 1,2-bis(2-phenylethynyl)benzene on Au(111) using scanning tunneling microscopy and computer simulations. Cyclization of sterically hindered enediynes is known to proceed via two competing mechanisms in solution: a classic $C^1$–$C^6$ or a $C^1$–$C^5$ cyclization pathway. On Au(111) we find that the $C^1$–$C^5$ cyclization is suppressed and that the $C^1$–$C^6$ cyclization yields a highly strained bicyclic olefin whose surface chemistry was hitherto unknown. The $C^1$–$C^6$ product self-assembles into discrete non-covalently bound dimers on the surface. The reaction mechanism and driving forces behind non-covalent association are discussed in light of density functional theory calculations.

**Introduction**

The cyclization of enediynes has become a relevant isomerization process in many disparate research fields. For example, enediyne cyclization is a bioactive process observed in natural antibiotics and is thus of great interest in drug-design and anti-cancer research.[1] It is also a valuable tool in materials science where it is used for the synthesis of extended conjugated polymers.[2] Regarding the latter, two basic strategies have been studied: a radical step growth polymerization of cyclized enediynes,[3,4] or a radical cyclization cascade along the backbone of a *poly*-(*ortho*-phenylene ethynylene) polymer by way of overlapping enediyne units.[5,6] A deep understanding of the microscopic mechanisms of these cyclization reactions would thus greatly promote their technological use in fields such as medicine, biochemistry, and nanotechnology.

Recently, a number of studies of enediyne cyclizations on surfaces have appeared. The synthesis of surface-supported conjugated molecular wires has been achieved.[7,8] Exploration of the reactions on surfaces has allowed their direct visualization at the single molecule level by scanning probe microscopy, providing key insight into reaction mechanisms.[9,10,11] Most of the systems studied, however, have been chemically complex and render a large number of different products.[8,9,10] Here, we report a study on the thermally induced cyclization of a simpler enediyne, namely 1,2-bis(2-phenylethynyl)benzene (**1**), and the subsequent non-covalent self-assembly of its cyclization products.

**Results and discussion**

Sublimation of a submonolayer coverage of **1** onto Au(111) held at 293 K renders a surface decorated with discrete molecules as shown in Figure 1a. A close-up image (Figure 2a) reveals a boomerang-shaped morphology reminiscent of the reactant's molecular structure. Density functional theory (DFT) simulations of the STM image of **1** on Au(111) (Figure 2b) are in good agreement with STM images of the as-deposited reactant and thus confirm that the molecules remain unchanged upon deposition onto Au(111) at 293K. The as-deposited reactants **1** adsorb preferentially onto the *fcc* sections of the Au(111) surface reconstruction and appear well-separated from each other in a correlated fashion (Figure 2d). This is supported by comparison of the experimental nearest neighbor distribution histogram along the *fcc* trenches and a calculated random one-dimensional distribution of impenetrable and non-interacting particles at the same coverage (Figure 2d).[12] The latter decays monotonically with increasing distance, while the peaked experimental distribution is indicative of a repulsive long-range interaction between monomers.[13] As coverage is increased, the nearest neighbor distance distribution maximum shifts towards lower values and molecules begin to adsorb in the *hcp* regions as well (Fig. S1).

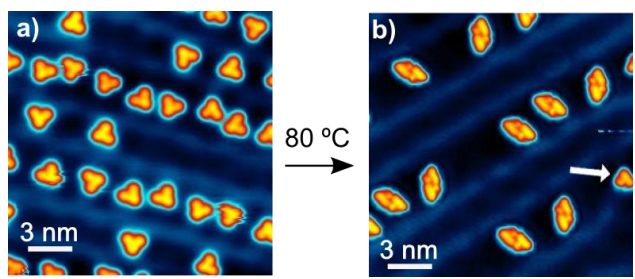

**Figure 1.** Experimental STM images of (a) the reactant (**1**) on Au(111) and (b) the products after annealing. The arrow in (b) marks a product monomer that is not found in a dimer. Imaging parameters are: $16 \times 16$ nm$^2$, $U = 0.1$ V, $I = 11$ pA.

The apparent repulsive intermolecular interaction may originate from electrostatic contributions. As previously observed with other electron donor molecules deposited on Au(111), repulsive Coulomb interactions can occur when molecules become charged on the substrate.[13] This scenario is corroborated by DFT calculations of **1** adsorbed on a Au(111) surface. The calculations show sizeable electron transfer from molecule to substrate which, according to a Bader analysis, amounts to 0.15 e⁻. The charge transfer is visualized in the isosurface contour plot of the differential electron density displayed in Figure 2e, and in the differential charge analysis, laterally-averaged over the xy plane, along the surface normal direction (Figure 2f).

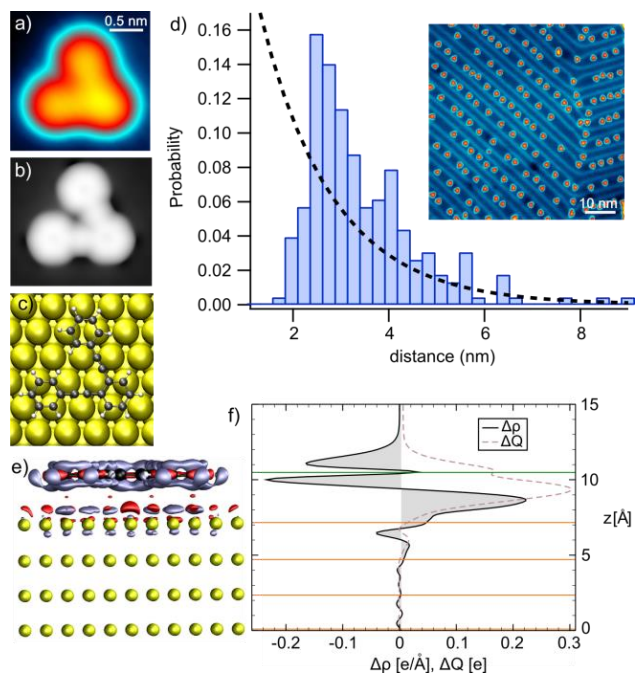

**Figure 2.** (a) Close-up experimental STM image of the reactant **1** (1.9 × 1.9 nm$^2$, $U$ = 0.1 V, $I$ = 11 pA), compared with (b) the associated STM image simulation based on (c) the relaxed structure of **1** on Au(111), where Au, C, and H atoms are represented by the yellow, black, and white spheres, respectively. (d) Nearest neighbor distance distribution histograms from the sample shown in the inset (histogram based on 227 data points). A random nearest neighbor distribution calculated for the same coverage is shown for comparison by the dashed black line (linear molecule density along *fcc* trenches = 0.22 nm$^{-1}$). (e) Iso-density contour plot of the differential electron density $\Delta\rho = \rho_{tot}-\rho_{slab}-\rho_{ads}$ for **1** on Au(111). The isodensity value is 0.003 e/Bohr$^3$. Red (blue) denotes positive/electron-rich (negative/electron-poor) regions. (f) Differential charge analysis laterally-averaged (over the xy plane) as a function of z of the valence electron density $\Delta\rho(z)$ and its integrated value Q(z). The mean z positions of Au layers and **1** are indicated by orange and green horizontal lines, respectively, and aligned with the corresponding contour plot in (e). There

is an accumulation of electrons (positive $\Delta Q$ or $\Delta\rho(z)$) at the interface between **1** and the Au(111) surface with a maximum at z~9 Å.

Upon annealing to $T$ = 353 K, a chemical transformation of **1** is induced, yielding a new surface morphology as shown in Figure 1b, where the majority of products associate into dimers. Only molecules adsorbed at the surface's reconstruction dislocations (herringbone kinks) and rare isolated molecules on the surface (as highlighted by an arrow in Fig. 1b) remain as product monomers, displaying similar STM contrast regardless of their location. Close-up images of the dimer aggregates and of the less frequent product monomers are shown in Figures 3j and 3a, respectively. The image contrast of the dimer complex corresponds to two monomers positioned side-by-side. This proves the non-covalent nature of the dimerization, since otherwise the new bonding structure and the associated changes in electronic properties would be reflected in a substantially different contrast in the STM images. Remarkably, virtually all reactant molecules transform into the same product structure, in sharp contrast to cyclization studies performed with the same precursor in solution [14,15] or with more complex precursors on surfaces.[8,9,10] Similar to the starting reactants, the dimerized products are seen to favor the *fcc* regions of the surface and to maintain a relatively large inter-dimer spacing (Fig. 4a). Their nearest neighbor distance distribution, peaked at 2.5 nm, clearly deviates from that of a random one-dimensional distribution at the same coverage (Fig. 4b) and is indicative again for repulsion between dimers.

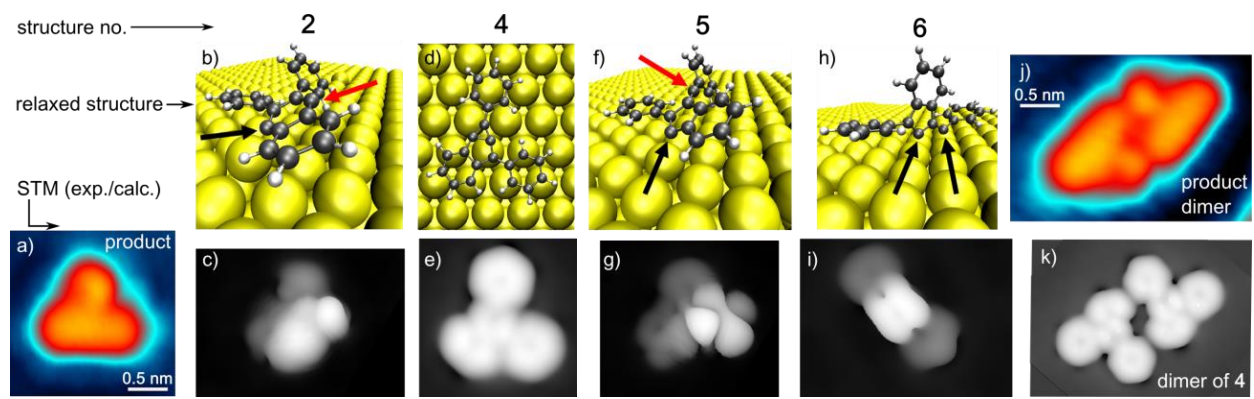

**Figure 3.** (a) Close-up view of an experimental STM image of a product monomer (1.9 × 1.9 nm$^2$, $U$ = 0.1 V, $I$ = 10 pA.). Relaxed structures and their associated STM image simulations are shown for **2** (b,c), **4** (d,e), **5** (f,g) and **6** (h,i) (structure **3** was not stable on the surface since it spontaneously transformed into **4**, according to calculations). Au, C, and H atoms are represented by the yellow, black, and white spheres, respectively. Radical carbon sites bound or not bound to Au are marked with black and red arrows, respectively. (j) Experimental STM image of the product dimer (1.9 × 2.7 nm$^2$, $U$ = 0.1 V, $I$ = 10 pA) for comparison with (k) the simulated STM image of a dimer of product **4**.

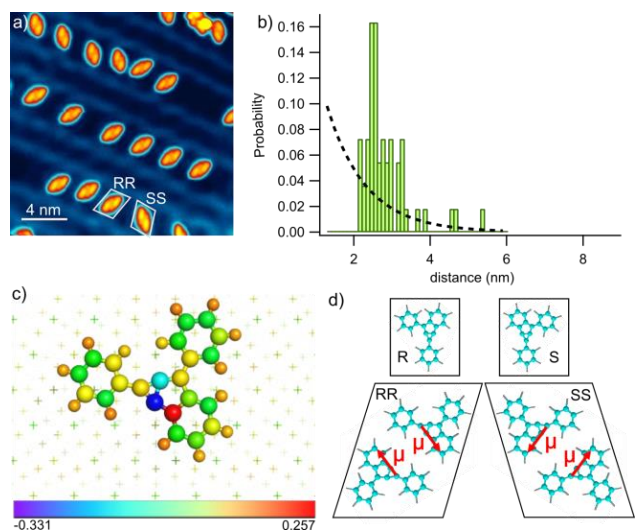

**Figure 4.** (a) Representative STM image of product dimers. Examples of dimers of RR and SS chirality are indicated and correspondingly labeled. (b) Nearest neighbor distance distribution histogram for dimers (based on 54 data points) compared to a random nearest neighbor distribution calculated for the same coverage (dashed black line, linear dimer density = 0.3 nm$^{-1}$). c) Bader

charge distribution for **4** on Au(111) reveals strongly polarized bonds within the bicyclic structure. The color scale marks the excess (blue) or deficiency (red) of electrons on each atom, see color scale bar. d) The charge distribution within **4** creates a strong in-plane molecular dipole moment μ (red arrows) that causes an attractive intermolecular interaction and drives the formation of antiparallel dimers. Monomers and dimers are shown with R, S, RR and SS chirality, respectively.

The possible transformations that reactant **1** may undergo upon thermal activation are displayed in Figure 5a. The dimerization through Glaser coupling of alkynes observed in previous works is prevented here by the terminal phenyl rings.[10,16] Instead, reactant **1** can undergo the following isomerization reactions. In a first step, $C^1$–$C^6$ (Bergman[17]) or $C^1$–$C^5$ (Schreiner-Pascal[14,18]) cyclization can lead to the biradical intermediates **2** and **5**, respectively.[14,18,19] For each of these two intermediates, a [1,2]-phenyl migration onto an adjacent $sp^2$ radical center may additionally take place, as readily observed in solution,[15] to alleviate steric repulsion and yield the stabilized products **3** or **6**. Finally, the biradical **3** may further undergo an isomerization to yield the strained benzannulated bicyclic diene **4**.[20,21]

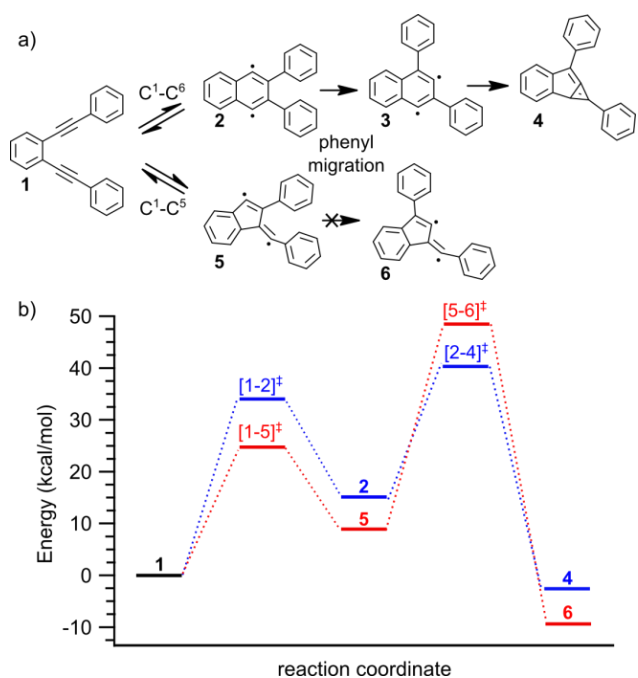

**Figure 5.** (a) Schematic representation of the precursor **1** and possible products after cyclization ($C^1$–$C^6$ or $C^1$–$C^5$) and subsequent phenyl migration processes. (b) Calculated relative energies (in kcal/mol) for all species and transition states adsorbed on Au(111). See Supporting Information for more details.

To assign chemical structures to the STM images, we have relaxed the geometries and simulated the STM contrast for all species on Au(111) using DFT calculations. The structure of the postulated intermediate **3** was not stable on the Au(111) surface; instead, **3** rapidly rearranges into **4**.[22] The results of the DFT calculations are displayed in Figure 3. Both **1** and **4** are physisorbed and lie flat on the Au(111) surface with an average adsorption height of ~3.3 Å with respect to the top layer of Au atoms. The other calculated species (**2**, **5**, **6**), however, are chemisorbed via the carbon-centered radicals with the free valences of the underlying substrate (C–Au bond distance ~2.14 Å), resulting in non-planar adsorbate geometries. These non-planar species are not assigned to the product due to the obvious mismatch between the simulated and experimental STM images.

However, the calculated STM contrast of **4** (Figure 3e) and its relaxed dimer (Figure 3k) very nicely reproduce the experiment, allowing assignment of **4** to the product's chemical structure. This assignment is further supported by the calculated Bader charge transfer of 0.21 e$^-$ from **4** to the surface, which is consistent with the observed repulsion between product dimers (Figure 4a).

The total energies of all species and barriers were calculated after relaxation both in gas phase and in the adsorbed state on a Au(111) surface. In the gas phase (see Supporting Information), all optimized species adopt a non-planar conformation to reduce the steric hindrance of adjacent phenyl rings. It is evident from the calculations that the $C^1$–$C^6$ cyclization mechanism is the dominant pathway in terms of activation barriers and stability of products in the gas phase, with **4** being 33.4 kcal/mol more stable than the $C^1$–$C^5$ cyclization product **6**. Indeed, it has been observed that the thermolysis of **1** in solution yields a $C^1$–$C^6$ to $C^1$–$C^5$ product distribution ratio of approximately 5:1.[15]

The results on Au(111) are summarized in Figure 5b. Interestingly, the computed stability is quite different and shows that our gas phase calculations have little bearing on the surface results. In particular, the $C^1$–$C^6$ cyclization on Au(111) faces an initial higher energy barrier than the $C^1$–$C^5$ pathway due to the steric hindrance of the approaching bulky terminal phenyl rings of the parent enediyne **1** and the surface-promoted planarity.[23] Furthermore, **6** now becomes the most energetically stabilized species due to its bidentate chemisorption mode on Au(111) (Figure 3h, black arrows). We note, however, that the activation energy barrier preceding the formation of **6** is very high (Fig. 5b) due to an interfering C–Au bond (see black arrow in Figure 3f). Thus, while **1** is in equilibrium with both **2** and **5**, i.e. **1** ⇌ **2** and **1** ⇌ **5**, only **2** can overcome the lower activation barrier leading to **4** at 353 K. Note that both phenyl migration steps (**2** to **4** and **5** to **6**) proceed via a spirocyclic transition state, similar to the one reported in solution/gas phase by Lewis and Matzger.[15] At the transition state, the "walking" phenyl ring adopts a perpendicular position to the plane of the molecule, resulting in a tilting of the entire adsorbate with respect to the Au (111) surface. However, the phenyl migration for the step from **2** to **4** is kinetically favored (Fig. 5b) as it does not require the breaking of any C–Au bond and proceeds via the free C radical in **2** that does not interact strongly with the surface (Figure 3b, red arrow). In contrast, the phenyl migration from **5** to **6** must proceed through a carbon-centered radical bound to the substrate (Figure 3f, black arrow)[24]. This results in a high barrier associated with the involvement of a C–Au bond.[25-27] At 353 K, intermediate **5** is then kinetically trapped and reverts to the starting material. Under our experimental conditions, product **4** is thus exclusively formed instead of the thermodynamically more stable product **6**.

The question remains as to what drives the dimerization in spite of the charged products and expected electrostatic repulsion.[28,29] Covalent bonding can be excluded because the product (**4**) quenches the initially generated biradical state through the formation of a benzannulated bicycle. Besides, the STM contrast for the dimer clearly resembles two neighboring monomers (Figure 1) rather than a completely hybridized new compound. According to DFT calculations, we estimate a relatively weak binding energy for the dimer of 1.6 kcal/mol on Au(111) and of 1.9 kcal/mol in the gas phase.[30] A significant part of the stabilization is attributed to attractive van der Waals (vdW) interactions. However, vdW alone would cause further polymerization (oligomers with n>2), which is not observed experimentally. Besides, vdW attraction is expected to be comparable for the reactant **1**, which does not undergo dimerization. Moreover, product **4** systematically arranges in a very specific anti-parallel alignment in the dimer, while the long-ranged vdW interactions are mostly indifferent to such geometric preferences. This suggests the existence of additional driving forces behind the dimerization.

A possible mechanism for the antiparallel arrangement in the dimer arises from the fact that DFT calculations predict that **4** has a strong electric dipole moment (2.19 D in gas phase and almost 3 D in the adsorbed geometry) due to the highly polarized bonds of the bicyclic structure. The product's intramolecular charge distribution is pictured in Fig. 4c for the adsorbed molecule, and shows how the charge transfer within each atom results in a strong in-plane molecular dipole moment. The opposite orientation of the monomers within the dimer thus arises from an attractive dipole-dipole interaction between the product molecules (Figure 4d).[31]

This dipole-dipole interaction explains the particular arrangement within the dimers, since it corresponds to the configuration in which the bicycles (and thus the dipoles) are closest and antiparallel to each other in order to maximize the attraction. A rudimentary electrostatics estimate shows that the stabilization energy due to the dipole-dipole interaction is a significant part of the total calculated binding energy. This estimate begins with the understanding that the interaction energy of a pair of dipoles is $\Delta E_{\mu-\mu} = \mu^2 cos\theta/(4\pi\varepsilon_0 r^3)$. Taking the dipole moment of **4** on the surface ($\mu$~3 D) and an angle $\theta$=180º for the antiparallel alignment, as well as $r$~6 Å for the distance between the most polarized bonds in the dimer, we obtain a total dipole interaction energy of $\Delta E_{\mu-\mu}$~0.6 kcal/mol, which is significant at the low temperatures of the STM measurement (T = 5 K).

This arrangement also implies that the molecules show chiral recognition upon dimerization (product **4** exhibits planar chirality on the surface, displayed in the insets of Fig. 4d). Although **4** is a racemic mixture of R and S enantiomers on the surface, dimers are always formed by molecules of the same chirality because it allows maximized intermolecular attraction (closest distance of the polarized bonds in an antiparallel alignment). Pairs of R or S chirality can be easily distinguished by their distinct azimuthal orientation within the reconstruction trenches, as marked in Figure 4a.

**Conclusions**

In summary, we report the cyclization reaction of 1,2-bis(2-phenylethynyl)benzene and subsequent dimerization of its cyclized products on Au(111). We find that the dominant reaction mechanism is a $C^1$–$C^6$ cyclization that leads to monomer **4**. Product **4** is a highly polarized bicycle with a large in-plane electric dipole moment that plays an important role in the subsequent association and antiparallel arrangement of non-covalent dimer complexes. This work also highlights the limitations of gas-phase calculations to predict reactivity on surfaces, in particular when intermediate species are chemisorbed.

ASSOCIATED CONTENT

**Supporting Information**. Experimental and computational details. Nearest neighbor distribution histogram of an additional sample with higher reactant coverage. Optimized structures in gas phase and on the surface. Full calculated energetics in gas phase and on Au (111). Computed potential energy diagram for the isomerization step from **3** to **4**. This material is available free of charge via the Internet at http://pubs.acs.org.

AUTHOR INFORMATION

**Corresponding Author**

*d_g_oteyza@ehu.eus

**Author Contributions**

‡ These authors contributed equally. The manuscript was written through contributions of all authors. All authors have given approval to the final version of the manuscript.

ACKNOWLEDGMENT

Research supported by the U.S. Department of Energy Office of Basic Energy Sciences Nanomachine Program under contract no. DE-AC02-05CH11231 (STM imaging), by the Office of Naval Research BRC Program (molecular synthesis), by the European Research Council grants ERC-2010-AdG-267374-DYNamo and ERC-2014-STG-635919-SURFINK (computational resources and surface analysis, respectively), by Spanish Grant no. FIS2013-46159-C3-1-P (simulated reaction landscape), and by Grupos Consolidados UPV/EHU del Gobierno Vasco no. IT-578-13 (simulated dimer binding energy). A.P.P. acknowledges postdoctoral fellowship support from "Ayuda para la Especialización de Personal Investigador del Vicerrectorado de Investigación de la UPV/EHU-2013" and from the Spanish "Juan de la Cierva-incorporación" program (IJCI-2014-20147). E. Goiri is acknowledged for help and discussion on the statistical analysis of interparticle distances.

[22] The isomerization between biradical monocyclic and closed-shell bicyclic structures of the closely related *meta*-benzyne molecule ($C_6H_4$) in gas phase has been extensively studied computationally. It was concluded that the monocyclic structure with equilibrium distance of 2.05 ± 0.05 Å between C radical centers was preferred and that the electronic structure is best represented as a σ–allylic C1C2C3 system.[21]

[23] At the transition state, the emerging C–C bond distance is 2.10 Å and 2.147 Å for $C^1$–$C^6$ and $C^1$–$C^5$, respectively. The resulting intermediates **2** and **5** do not show the expected zwitterionic charge distribution because one of the radical centers is quenched on the Au (111) surface.

[24] In the phenyl shift of step **5** to **6**, the disappearing $sp^2$ C radical in **5** (black arrow in Fig.3f) is lifted off the surface, while the newly emerging C radical center in **6** sinks and binds to the Au(111); at the transition state, however, both C centers are bound to the Au (111) surface.

[25] The Bell-Evans-Polanyi principle does not apply here because of the drastically different involvement of the surfaces.